\documentclass[12pt,preprint]{aastex6}

\begin{document}

\def \lsun          {\hbox{L$_{\odot}$}}
\def \msun          {\hbox{M$_\odot$}}
\def \rsun          {\hbox{R$_\odot$}}

\title{Coevality in Young Eclipsing Binaries}

\author{M.Simon\altaffilmark{1,2} \& Jayashree Toraskar\altaffilmark{2}} 

\altaffiltext{1}{Dept. of Physics and Astronomy, Stony Brook University, Stony Brook, NY 11794-3800, USA; michal.simon@stonybrook.edu}
\altaffiltext{2}{Dept. of Astrophysics, American Museum of Natural History, New York, NY 10024, USA; toraskar@amnh.org}

\begin{abstract}

The ages of the components in very short period pre-main sequence (PMS) 
binaries are essential to an understanding of their formation.
We considered a sample of 7 PMS eclipsing binaries (EBs)  with
ages 1 to 6.3 MY and component masses 0.2 to 1.4 \msun.  The very high 
precision with which their masses and radii have been measured, and
the capability provided by the 
{\it Modules for Experiments in Stellar Astrophysics (MESA)} 
to calculate their evolutionary tracks at exactly 
the measured masses, allows the determination of  age differences  of the 
components independent of their luminosities and effective temperatures.
We found that the components  of 5  EBs, ASAS J052821+0338.5, Parenago 1802,
JW 380, CoRoT 223992193, \& UScoCTIO 5, formed within 0.3 MY of each other. 
The parameters for the components of V1174 Ori, imply an implausible 
large age difference of 2.7 MY and should be reconsidered. The 7th EB in
our sample, RX J0529.4+0041 fell outside the applicability of our analysis.

\end{abstract}


\section{Introduction}

The components of binaries are thought to form together and the differences
in their ages are usually assessed  by placing the components on a 
Hertzsprung-Russel Diagram (HRD).  Using this 
approach \citet{2009ApJ...704..531K} 
showed that angularly resolvable binaries in the Taurus star-forming 
region (SFR) are coeval to $\sim 0.16$ dex corresponding to an age difference 
of about 0.9 MY at age 2 MY. Eclipsing binaries (EBs) provide precise
 measurements of their component masses, radii, effective temperatures, 
and luminosities.  These parameters are fundamental and provide inputs 
for tests of theoretical models of stellar evolution (e.g. 
\citet{2010A&Rv..18...67T}). Stellar masses and radii are the 
parameters most directly determined by the observations, are
model-independent, and can be measured with exquisite precision, sometimes 
better than  $1\%$.  In this paper we use  measured masses and radii 
of pre-main sequence (PMS) EBs to set limits on the age differences 
of their components that are smaller than previously available.

The radius of a PMS star decreases as it  collapses to the main sequence
along an evolutionary track that is determined by its mass.  We calculate 
evolutionary tracks of EB components at their  measured masses using
the {\it Modules for Experiments in Stellar Astrophysics (MESA)} 
(\citet{2015ApJS..220...15P} and earlier references therein) and 
determine their ages with reference to the measured radii.
 We do not  regard the ages we derive as absolute
because ages derived using other theoretical models could differ.  
Nonetheless we show that the age differences of the components we derive 
using the {\it MESA} tracks provide an accurate and precise  assessment 
of the components' age differences. \S 2 presents our sample of EBs and
describes the determination of their masses and radii.
\S 3 describes our analysis and its limitations and \S 4 presents our results.
\S 5 discusses the 5 EBs we find to be coeval to 0.3 MY on average and the
problematic age difference of the components of V1174 Ori.
 We summarize our  results in \S 6  and 
suggest future work.

\section{EB Sample}

EBs are rare and the number of known PMS EBs is small.  
\citet{2014NewAR..60....1S} presented a list of 13 well-studied PMS EBs 
with masses 4 \msun ~or less.  Here we study only EBs with masses in the 
mass range $\sim 0.2$ to 1.4 \msun ~because, as we discuss in \S 3, that 
is the range in which
we validated our use of evolutionary tracks calculated using {\it MESA}.
Table 1 lists the 7 EBs in our sample.  Six are from the compilation
by \citet{2014NewAR..60....1S} and one, UScoCTIO 5 was discovered in the 
 K2 Mission (\citet{2015ApJ...807....3K}; \citet{2016ApJ...816...21D}). 
The first 2 columns list the EB name, 
references for the parameters used, and the component to which the values
in cols 3-7 apply. Component masses and radii are the only parameters 
we need  in this work. Cols. 3, 4, and 5 list the  measured orbital 
period (days),  component masses, $M/\msun$, and photospheric radii, 
$R_p/ \rsun$.  The masses and radii are listed at the full precision 
derivable from the observations but the period is listed at far lower 
precision because its precise value
is unimportant here. We list the parameters as given by 
\citet{2014NewAR..60....1S} rather than the values in the original discovery
papers because Stassun et al. made critical assessments of 
their precisions. Col. 1 also lists
the EB's discovery paper or one that originally presented its parameters.
For UScoCTIO 5 we list the 
parameters derived by \citet{2015ApJ...807....3K} and 
\citet{2016ApJ...816...21D}. We analyze both to explore differences in
an assessment of the components' coevality.

All the EBs in our sample are short period double-lined 
spectroscopic binaries with well-studied light curves.
The orbital periods can be determined from either the 
photometric or spectroscopic data.  Parenago 1802 is 
the youngest EB in our 
sample.  With P=4.67d and eccentricity $e \sim0.02$, it has 
not yet circularized its orbit \citep{2012ApJ...745...58G}. 
UScoCTIO 5 has the longest period, P=34d,
and the largest eccentricity $e \sim 0.27$ 
(\citet {2015ApJ...807....3K}; \citet{2016ApJ...816...21D} 
The other EBs have periods 
less than $\sim 5.3d$ and circular orbits.  The mass
ratios of the components are in the range 0.58 to nearly
1.0.

Masses are derived from 
an analysis of spectroscopically measured radial 
velocities, RVs.  The RVs provide the velocity amplitudes, 
$K_A$ and ~$K_B$, eccentricity $e$, and the systemic 
velocity $\gamma$.  The velocity amplitudes, 
$K_{A,B}$, yield  the mass ratio directly, and, with
$P$~and $e$, the values of $M_{A,B}sin^3 i$.  
The  photospheric radii and inclination, $i$, 
are determined from analysis of the eclipses in 
the light curve. The ratio of the radii $R_A/R_B$ 
is derived from timing in the eclipses. Using 
also the $K_{A,B}$ the radii measured relative to 
the orbital semi-major axis are derived. 
Determination of these parameters is complex because 
it needs inputs that describe the radiation 
of the components, particularly their flux ratio.
The well-established Wilson-Devinney (WD) code 
\citep{1971ApJ...166..605W} and its descendants 
(e.g. PHOEBE, \citet{2005ApJ...628..426P}) are often 
used for this analysis. In the sample we are considering
JKTEBOP which needs fewer stellar inputs,
\citep{2007A&A...467.1215S} and \citep{1981AJ.....86..102P},
was used to analyze the light curves of JW380 \citep{2007MNRAS.380.541I},
CoRoT \citep{2014A&A...562A..50G},
and UScoCTIO 5 \citep{2016ApJ...816...21D}.
When analysis of the light curves reveals 
a signal in the eclipse minima that cannot be attributed
to the components, it is described as the ``Third Light''
parameter. Third light can arise in the starlight of a tertiary 
or of an unrelated star in the photometric aperture.  Third light
masquerading as a  signal can also result from the data processing.
Third light has been identified in four of the EBs in our 
sample: RX J0529.4+0041, V1174 Ori, Parenago 1802, and 
JW380 \citep{2014NewAR..60....1S}.

\section{Method of Analysis and its Limitations}
 
We use the the {\it MESA} code to calculate evolutionary tracks
because it provides the capability to calculate evolutionary
tracks at exactly the measured masses.  Thus it avoids the necessity
to interpolate between the masses at which published theoretical 
models (e.g. \citet
{2015A&A...577A..42B}, BHAC15; \citet{2016A&A...593A...99F}, F16)
are available. 
Appendix A provides more details about our use of the  code.
By referring to the dependence of the photospheric radius $R_p$~on age 
for the particular mass track,  we use the measured radius and its 
$\pm 1 \sigma$~values to determine the component's age and uncertainty.

We compared the tracks calculated by {\it MESA} to those of 
BHAC15 and F16.  Fig. 1  shows 
the $R_p~vs~log(Age)$ dependence of the  {\it MESA}, BHAC15, 
and F16 models for stars of mass $0.1 - 1.4 \msun$ over 
the range 1 to 100 MY.  The tracks provided by the {\it MESA}, 
BHAC15, and F16 models are very similar for stars older than 1 
MY and more massive than $\sim 0.5 \msun$.  The BHAC15 and 
{\it MESA} tracks agree down to $\sim 0.2 \msun$ and
age greater than 1 MY. Therefore we limited our sample of EBs analyzed
to 0.2-1.4 \msun.

\begin{figure}
\centering
\figurenum{1}
\includegraphics[width=7cm]{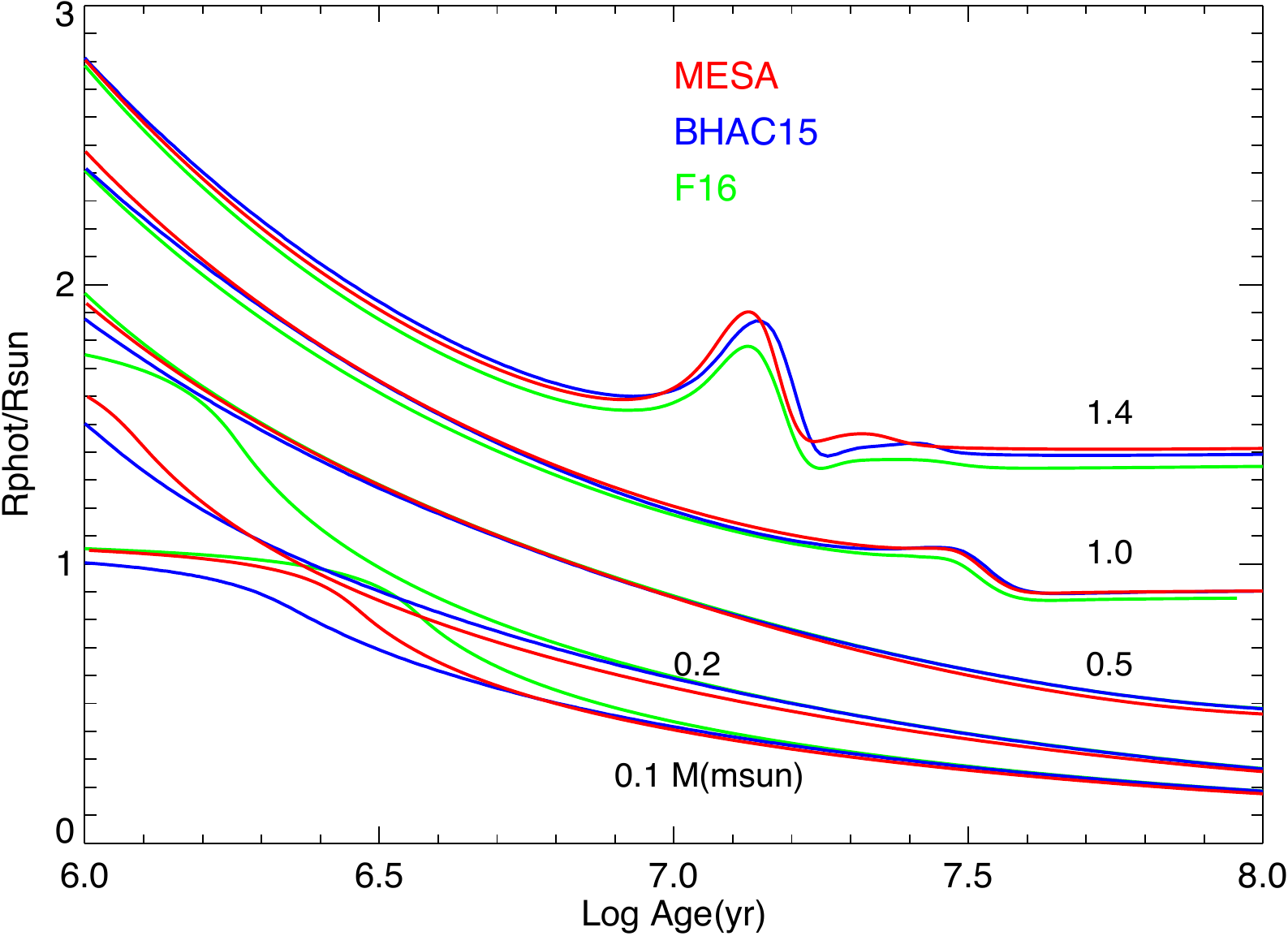}
\caption{A comparison of the $R~vs~log(Age)$ evolutionary tracks for the
{\it MESA}, BHAC15, and F16 (non-magnetic) models.}
\end{figure}

The effects of energy production by the CN cycle manifest
themselves as a ``puffing up'' of the radius at $\sim 30$ MY 
in 1 \msun~ stars and earlier for more  massive stars  
(Baraffe, priv. comm.; Feiden, priv. comm. and footnote 1 in 
\citet{2014NewAR..60....1S}).  This marks the transition from
the star's evolution along the Hayashi track  to the
Henyey track.  The figure shows that these effects do 
not appear in stars less massive than $\sim 0.5$ \msun.
This illustrates one limit on the applicability
of this age determination technique: For stars
more massive than $\sim 1$ \msun ~the  ``puffing up''
becomes a problem at ages greater than $\sim 10$ MY.

Fig. 1 indicates that precision reached in the
measurement of the age difference of the 
components depends on the slope of the 
$R_p~vs~log(Age)$ track. We can quantify
this easily.  For an uncertainty 
$\Delta R_p$ in the  $R_p$ measurement, 
the precision of the age estimate, 
$\Delta log(\tau)$, where $\tau$ is $log(Age)$, 
depends on the slope of as 
$$\Delta log(\tau) \sim {\Delta R_p \over {|dR_p/d\tau|}}.$$
The relation is approximate because all $R_p~vs~log (Age)$~
tracks have some curvature.  The slope of the track
for a $0.5 \msun$~star at $log(Age)$= 6.5, at which
$R_p = 1.3 \rsun$, is about -0.95 \rsun/MY.  A $2\%$ 
relative uncertainty in the measurement of $R_p$
is representative of the values in Table 1. The
relation for an estimate of the age uncertainty
yields $\Delta log(\tau) = 0.026$ dex for the
star in this example, comparable to the 
uncertainties presented in \S 4 and listed in Table 1 Col. 6.
As the contraction of the stars slows, the slopes
of the $R_p~vs~log(Age)$ tracks flatten; the 
smaller the slope the greater the uncertainty 
on the age.  Thus the technique is not useful at
ages greater than $\sim 10$ MY either because
of the operation of the CN cycle or because
of the slowing contraction of the stars or both.
At masses $\le \sim 0.2 \msun$~and ages $\le \sim
6.6$ dex, the models differ in their rates of
decrease of $R_p vs log(Age)$, probably because
of differences in their treatment of deuterium 
burning.  The figure suggests that the ``sweet spot'' 
for the application of this technique is a region
approximately centered on  $log(Age)=6.5$, extending
in age $\pm 0.5$ dex and over the mass range
$\sim 0.2 - 1.4 \msun$.  The similarity of rates 
of contraction indicates that the ages, their 
differences, and uncertainties determined 
using {\it MESA} tracks would be consistent 
with values obtained using the BHAC15 and F16 
models at the same values.  We discuss in 
\S 5 the effects of including magnetic fields.

\section{Results}

Fig. 2 shows the analysis of UScoCTIO 5 
in the $R_p/\rsun-log (Age)$ plane using 
\citet{2015ApJ...807....3K} and \citet{2016ApJ...816...21D})'s
parameters. The solid horizontal lines 
indicate the measured values of the radii of the A and B 
components; the dashed horizontal lines lie at 
$R_p/\rsun~\pm~1 \sigma$. The solid diagonal lines 
plot the $R_p~vs~Age$ evolutionary tracks and the dashed lines 
plot the tracks for $M_*$ and $M_*\pm 1\sigma$.  The component 
age is given  by the intersection of the 2 solid lines.  The
\begin{figure}
\centering
\figurenum{2}
\plottwo{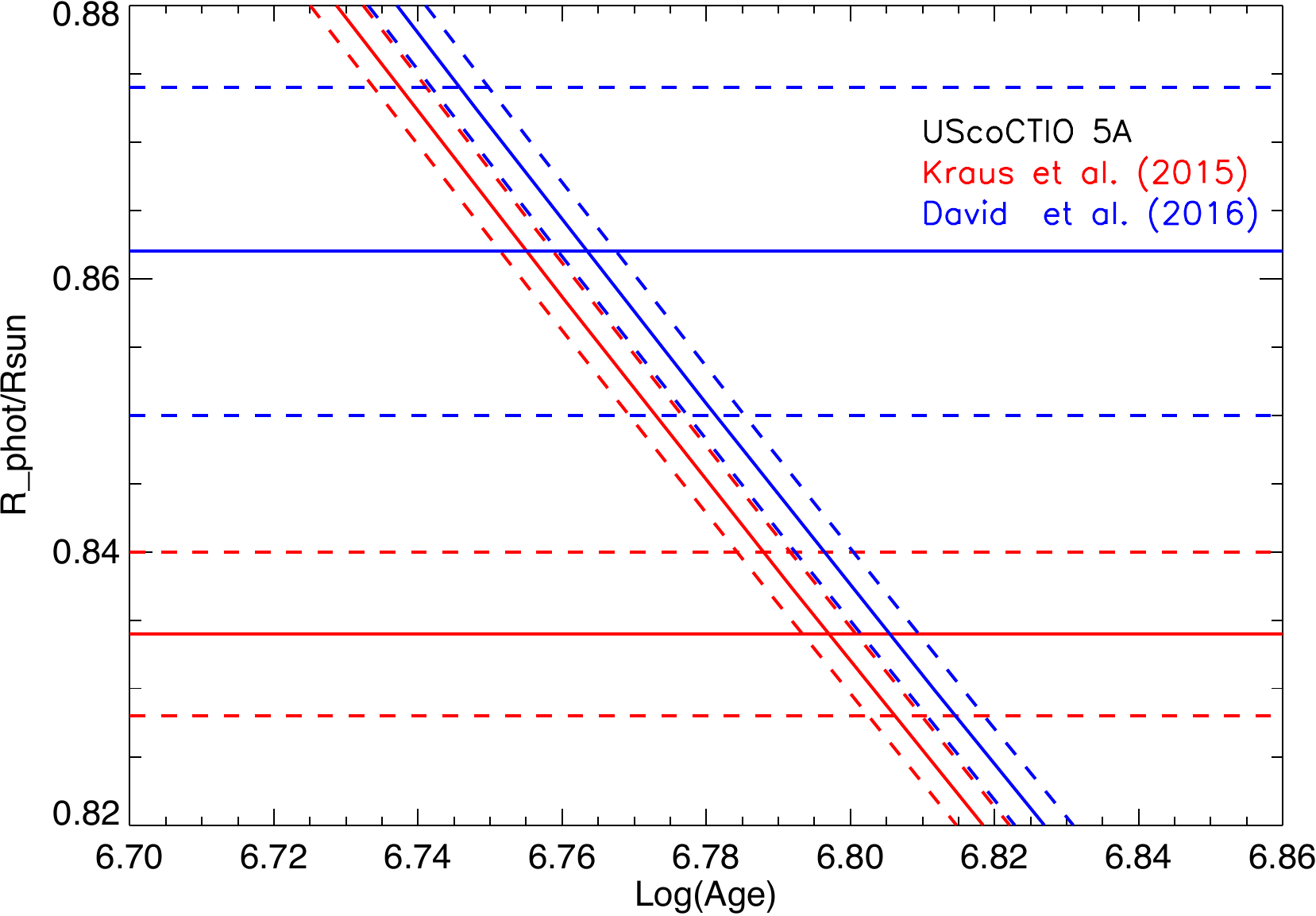}{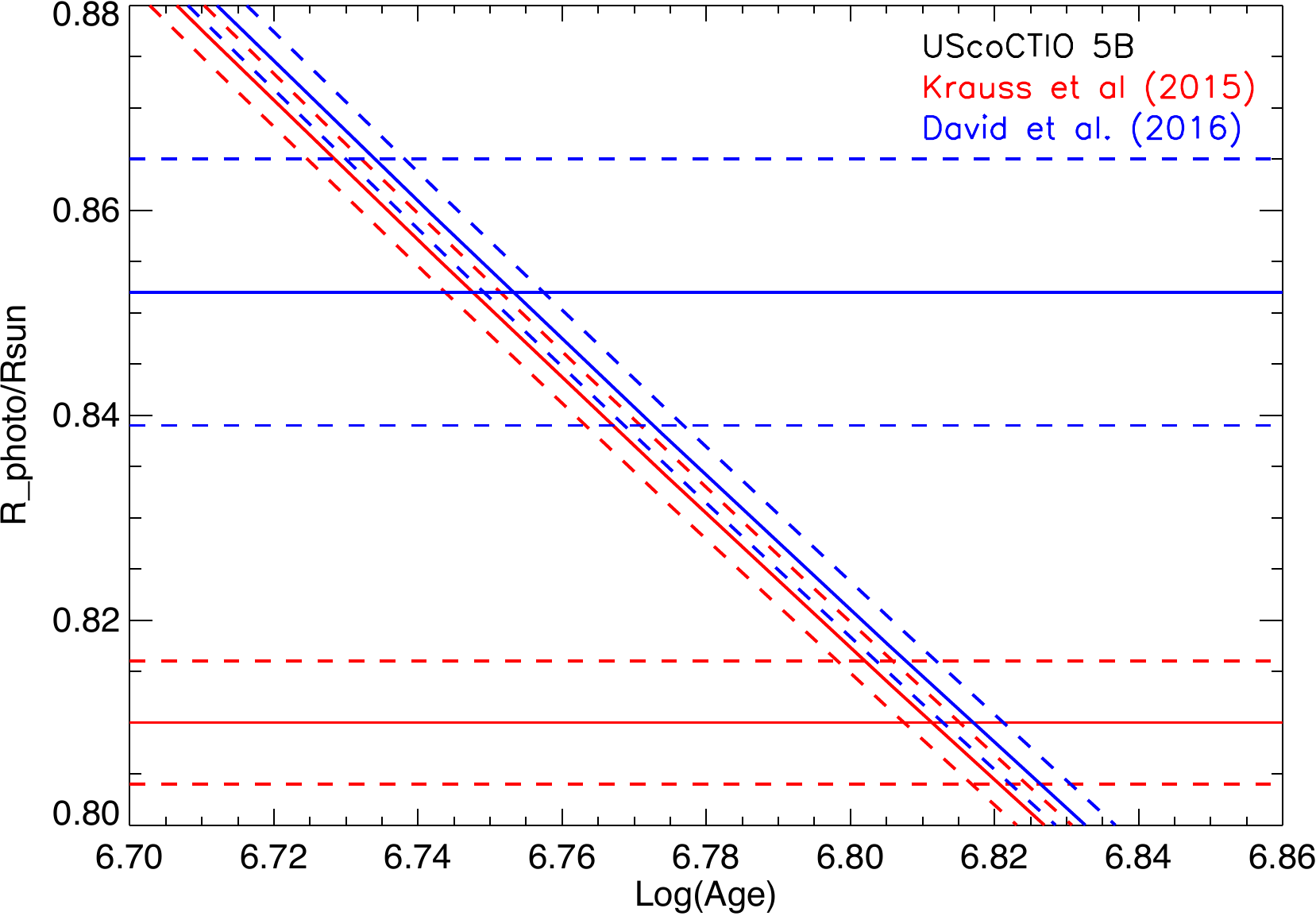} 
\caption{Analysis of the A and B Components of UScoCTIO 5 using the 
parameters derived by Kraus et al. (2015) and by David et al. (2016)
in {\it red} and {\it blue} respectively.  The age differences, 
(A-B), are  $-0.19\pm0.22$~MY. for the Kraus et al. parameters
and $+0.12\pm0.34$ and for David et al.'s. The two values do not
differ at a statistically significant level. }
\end{figure} 
locus of the $1 \sigma$~age uncertainty is described by an oval 
in the $R_p-log (Age)$ plane. It is 
unnecesssary to derive the equation of the oval because
its extrema at $(R_p/\rsun -  1\sigma)$ and $(R_p/\rsun + 1\sigma)$
give the $\pm 1 \sigma$ age uncertinty.  We list the component age and 
uncertainty 
in Col. 6 and age difference in Col. 7.  Even though the values 
of the B component's 
radius  differ by nearly $3\sigma$ the effect on an assessment 
of coevality is nil.  Neither result suggests a significant age 
difference and we can be confident that the ages of the components
are within 0.34 MY of each other.

Fig. 3 shows the application of this technique to RX J0529.4+0041
and demonstrates the limits of its usefulness.
Again, the straight lines show the measured photospheric radii
and their $\pm 1\sigma$ uncertainties.  The  $R_p~vs~log~Age$~tracks 
depart sufficiently from a linear dependence (Fig. 3a) that 
\begin{figure}
\centering
\figurenum{3}
\gridline{\fig{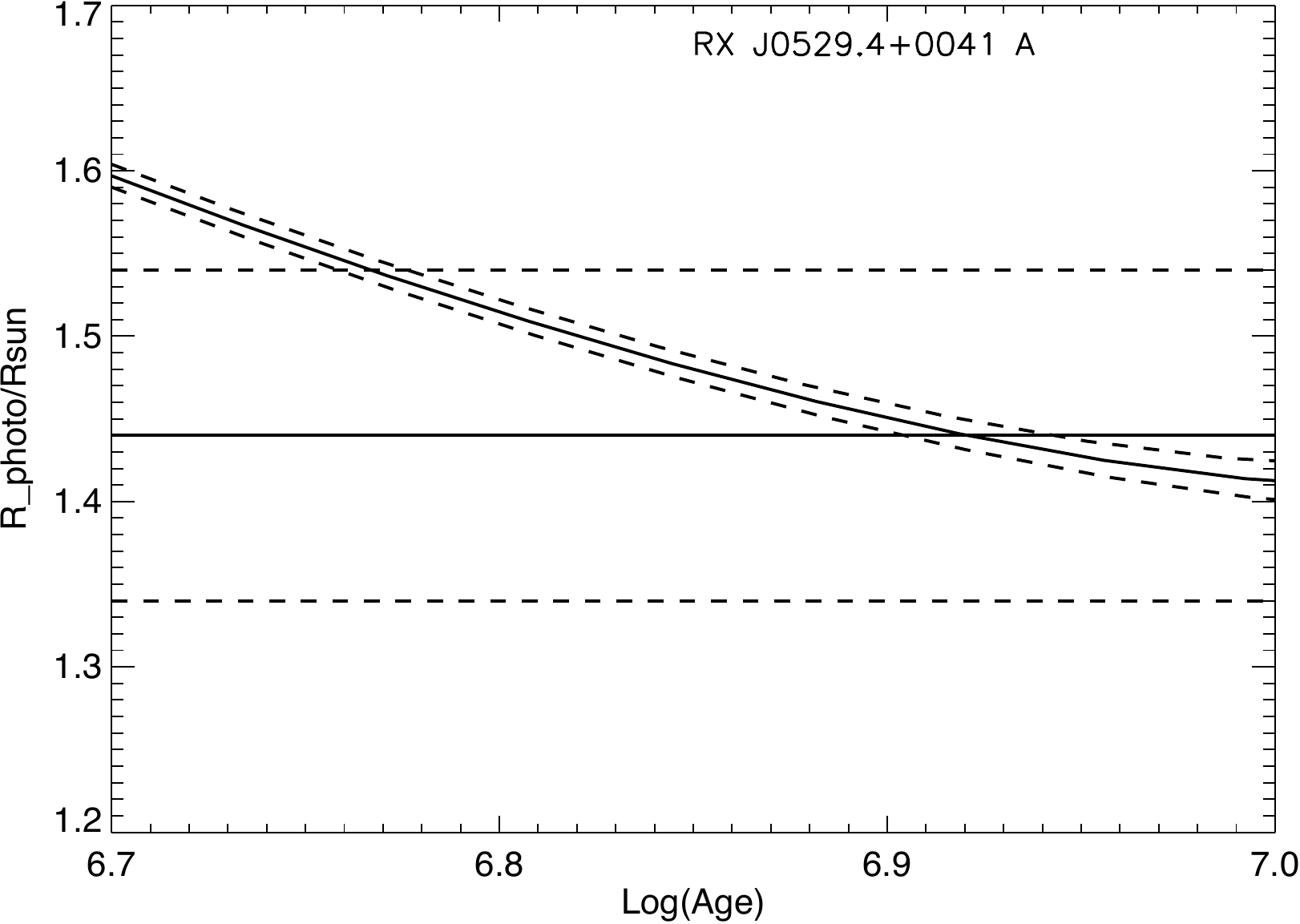}{0.3\textwidth}{(a)}
          \fig{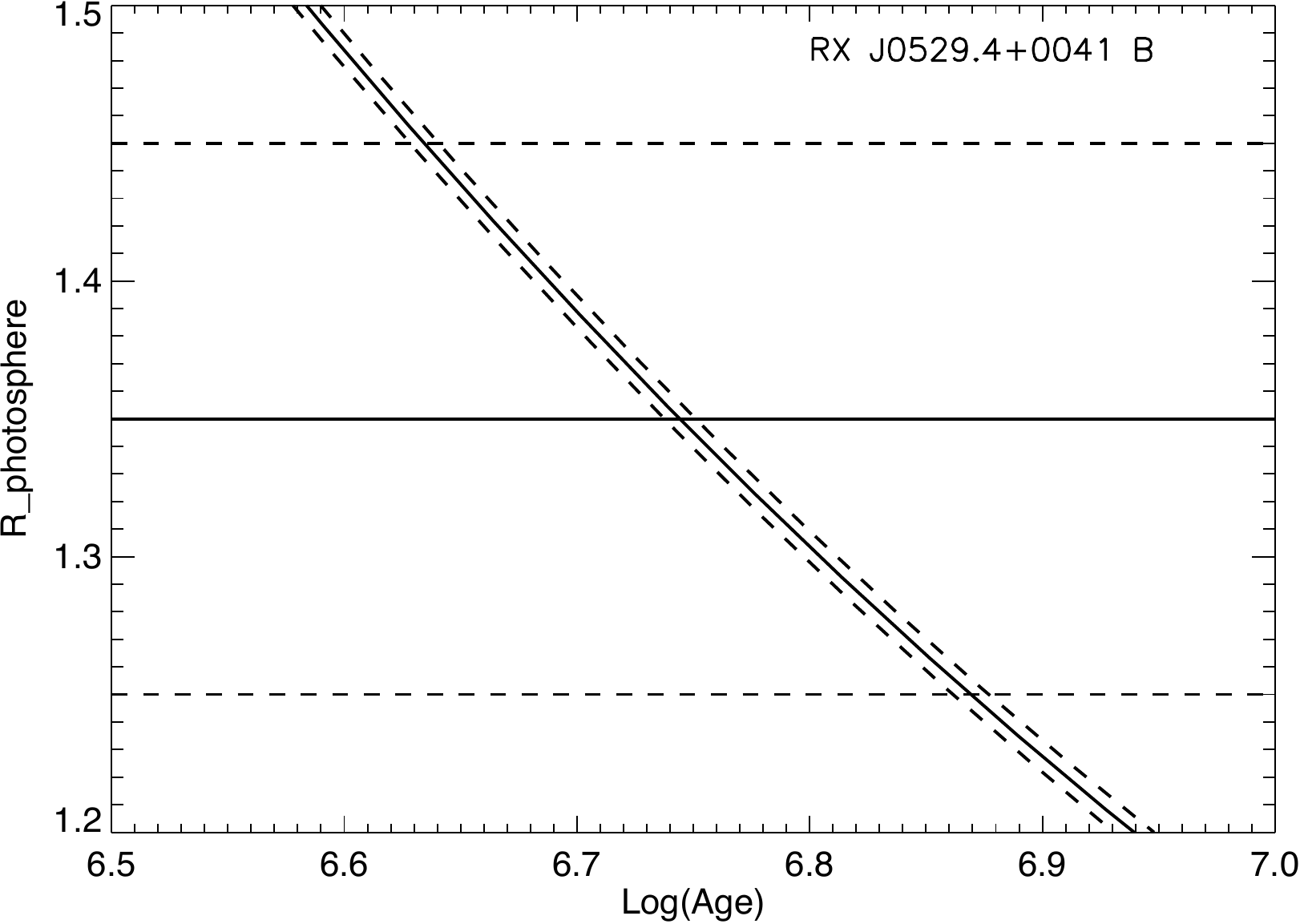}{0.3\textwidth}{(b)}
          \fig{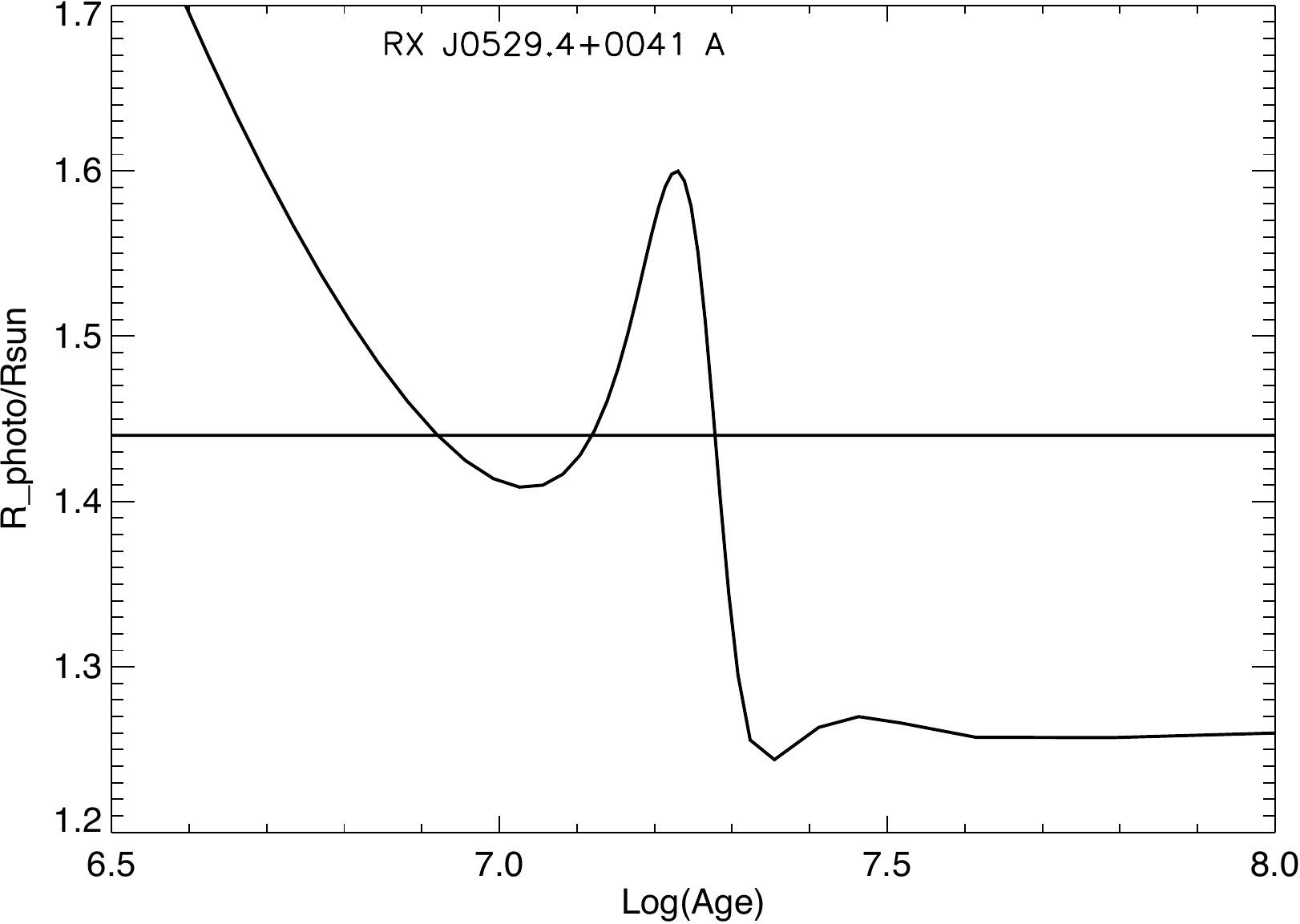}{0.3\textwidth}{(c)}
          }
\caption{Analysis of the Components of RX J0529.4+0041, 
A (Panel a) and B (b).  Panel (c) replots the A component 
analysis on a scale 
6.5 to 8.0 MY dex.  The brief increase in the star's radius at 
Log(age) $\sim 7.2$ is the result of CN burning (see text).  }
\end{figure}
it  is not possible to specify a unique $1\sigma$ upper bound on the 
A component's age. To provide a wider perspective,
Fig. 3c  plots the $R_p~vs~Age$ tracks of the A component
over the $log(Age)$ range 6.5 to 8.0.  
It shows that the  ``puffing up'' in radius discussed in \S 3, here
at about 16 MY, causes the age estimated by our
approach to be multivalued.

Fig.4  shows the  analysis applied to ASAS J05281+0338.5.
Although the masses are close to 1.4 \msun ~(Table 1) 
the figures show that the ages of the A and B components are 
young enough, $\sim 3.6$MY, that the curvature of the $R_p~vs$ 
log(Age) is small and does not compromise estimates of their
age and age uncertainties. The difference in ages of the
components in the (A-B) sense is -$(0.53\pm 0.85)$MY. 
Their ages are probably within 0.85 MY of each other.
\begin{figure}
\centering
\figurenum{4}
\plottwo{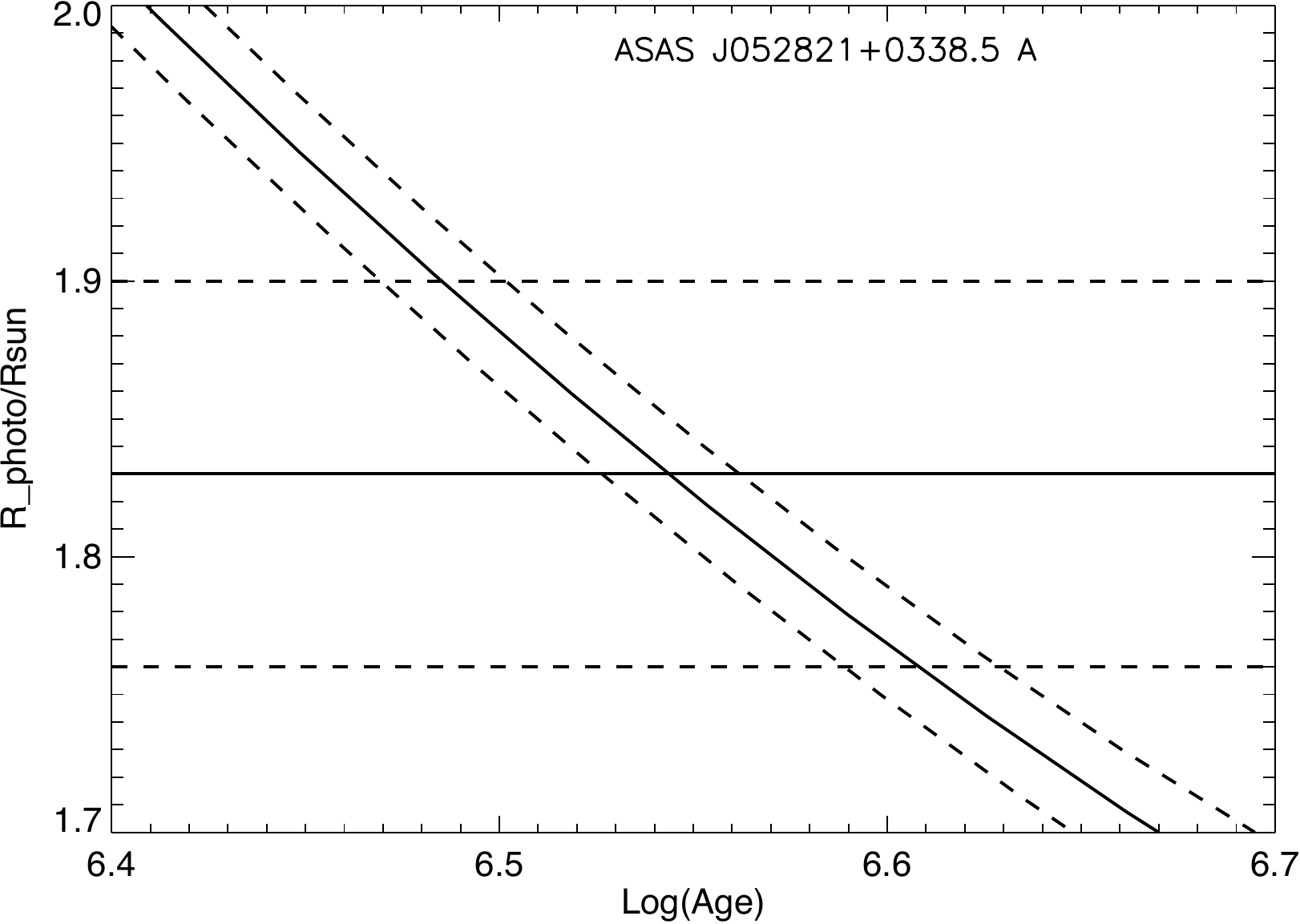}{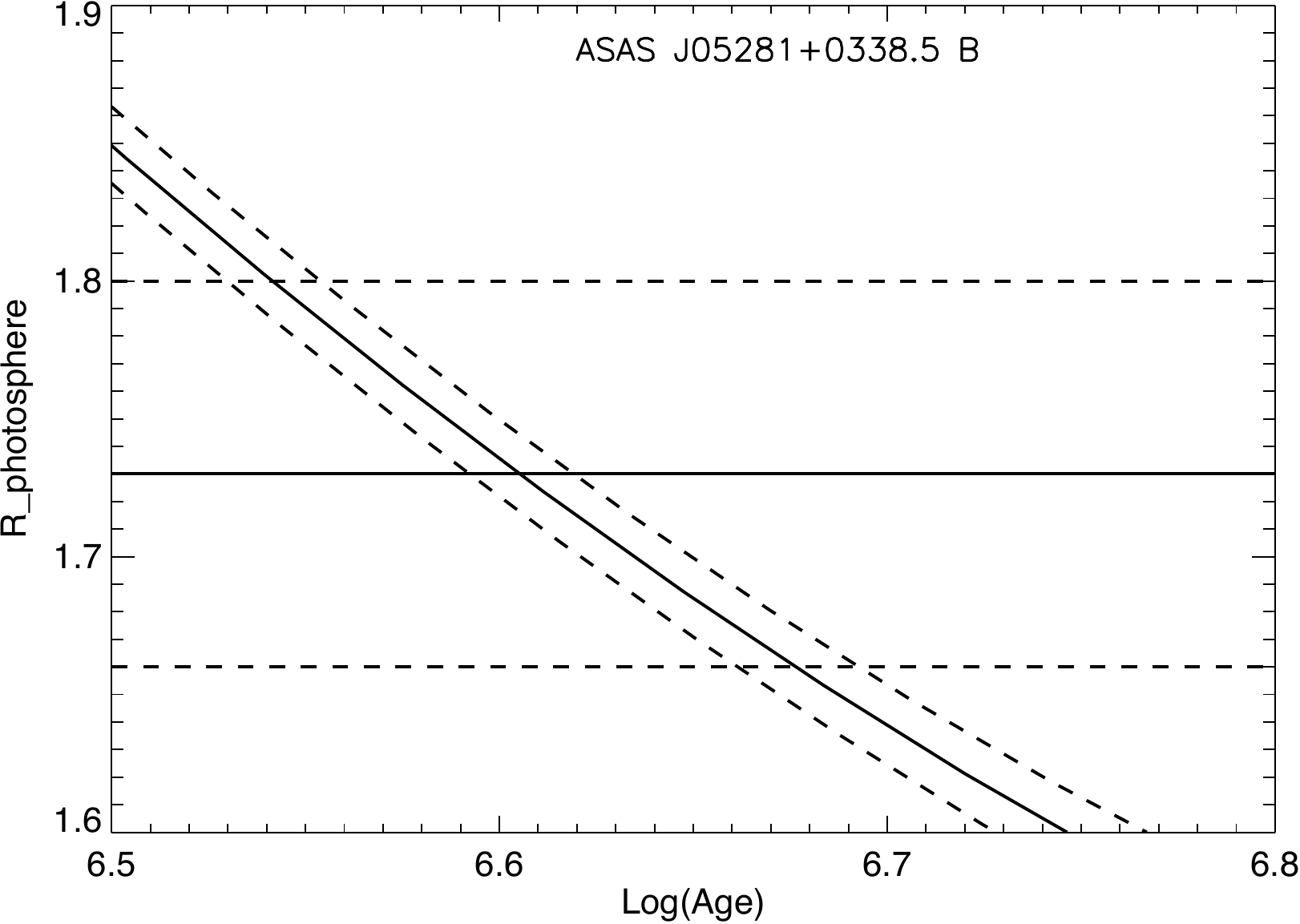}
\caption{Same analysis as in Fig.2 but for ASAS J05281+0338.5.  
The (A-B) age difference is $-0.53\pm0.85$ MY.}
\end{figure}

\begin{figure}
\centering
\figurenum{5}
\plottwo{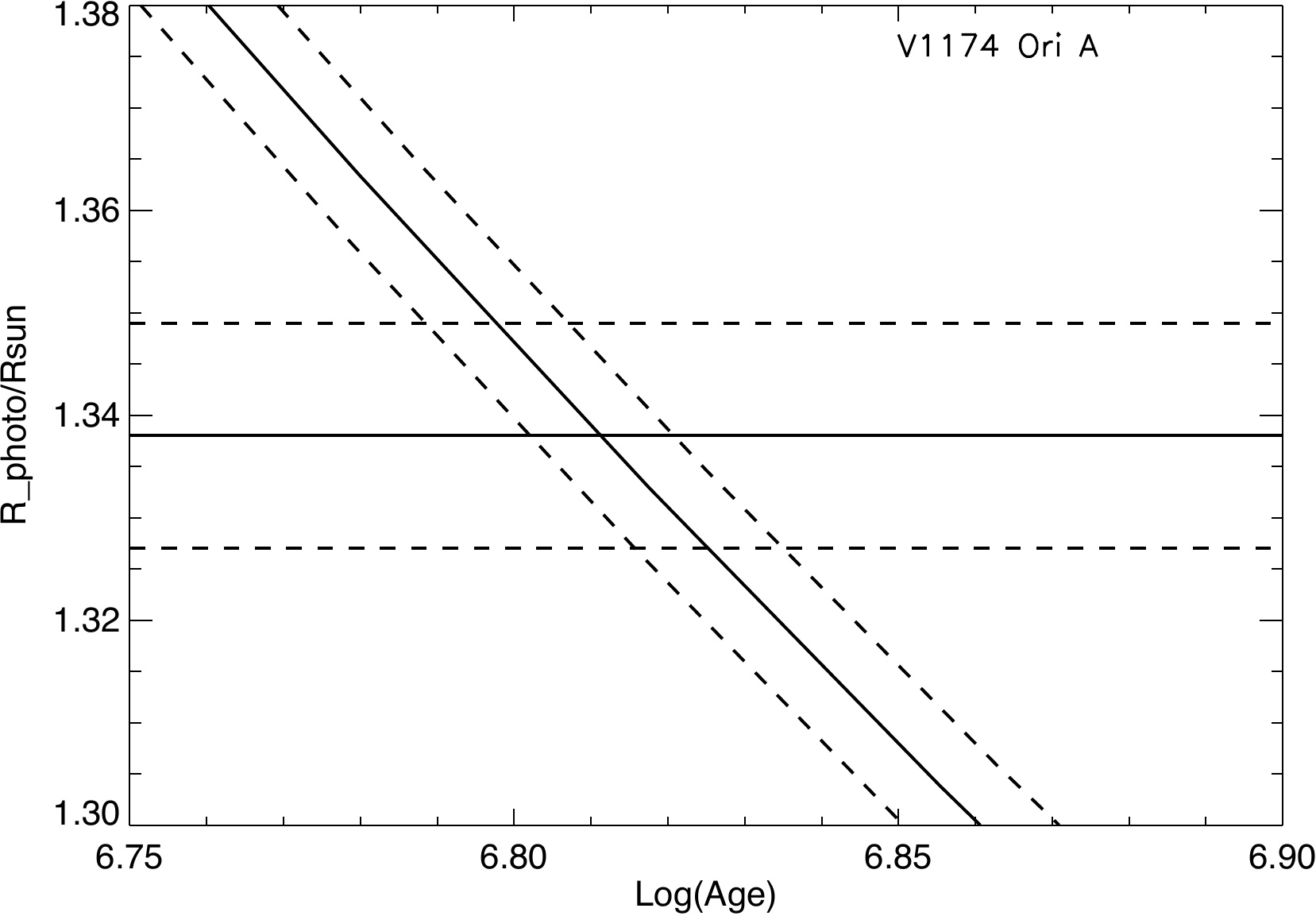}{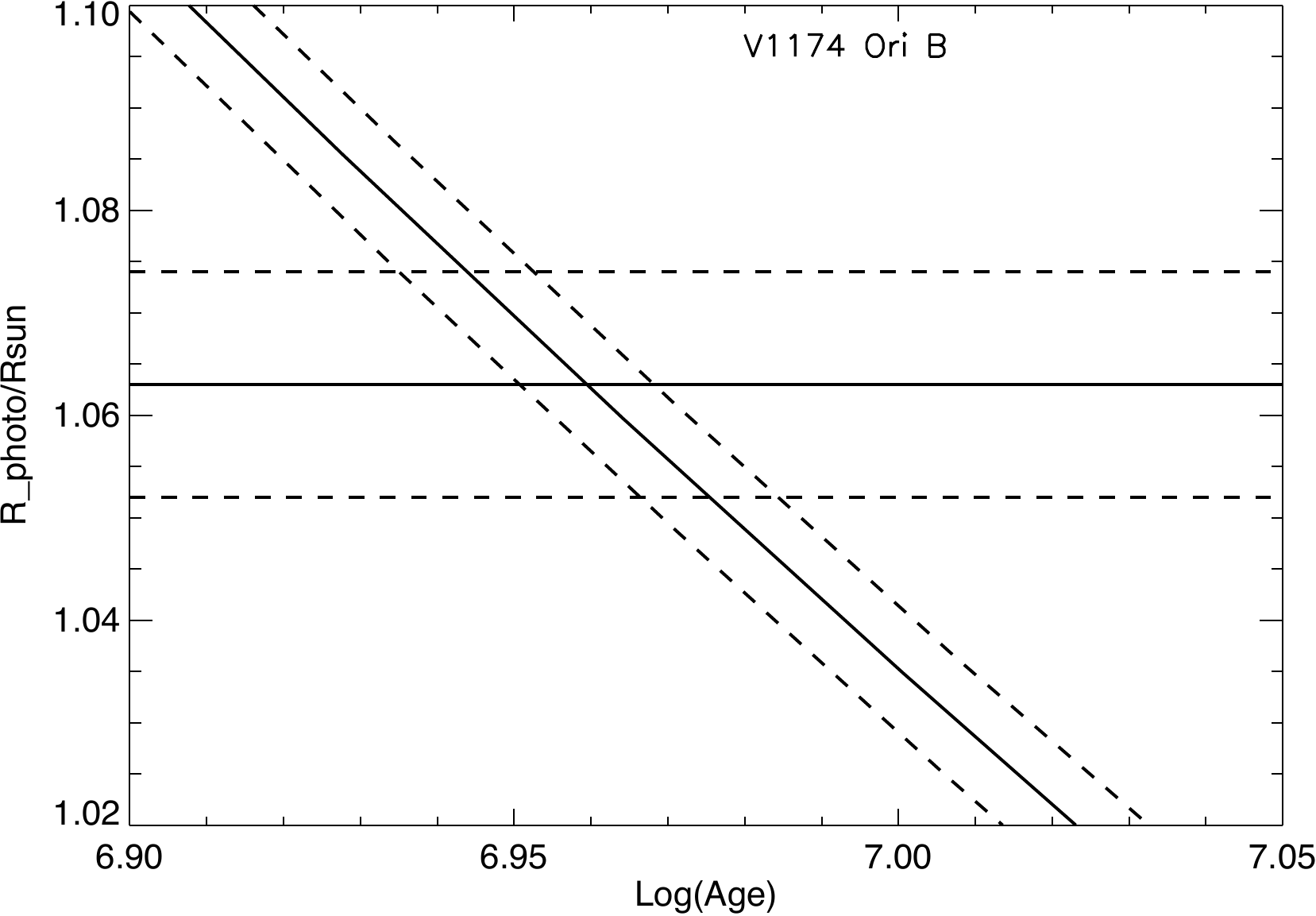}
\caption{Same as Fig. 2 but for V1174 Ori.  We regard 
the (A-B) age difference
as subject to confirmation, see \S 5.1.}
\end{figure}
Figs. 5, 6, 7, and 8 show the results of this analysis
for V1174 Ori, Parenago 1802, JW 380, and CoRoT 223992193.
Of our sample of 7 EBs,  V1174 Ori is the only EB  
whose components have, at least formally, a statistically 
significant age difference. The apparent age difference  
is such that it would make the B component 2.7 MY
older than A.   It is important to 
identify the reasons for the finding and to assess whether
the result is misleading.  We discuss our results and 
V1174 Ori in particular in the next section.

\begin{figure}
\centering
\figurenum{6}
\plottwo{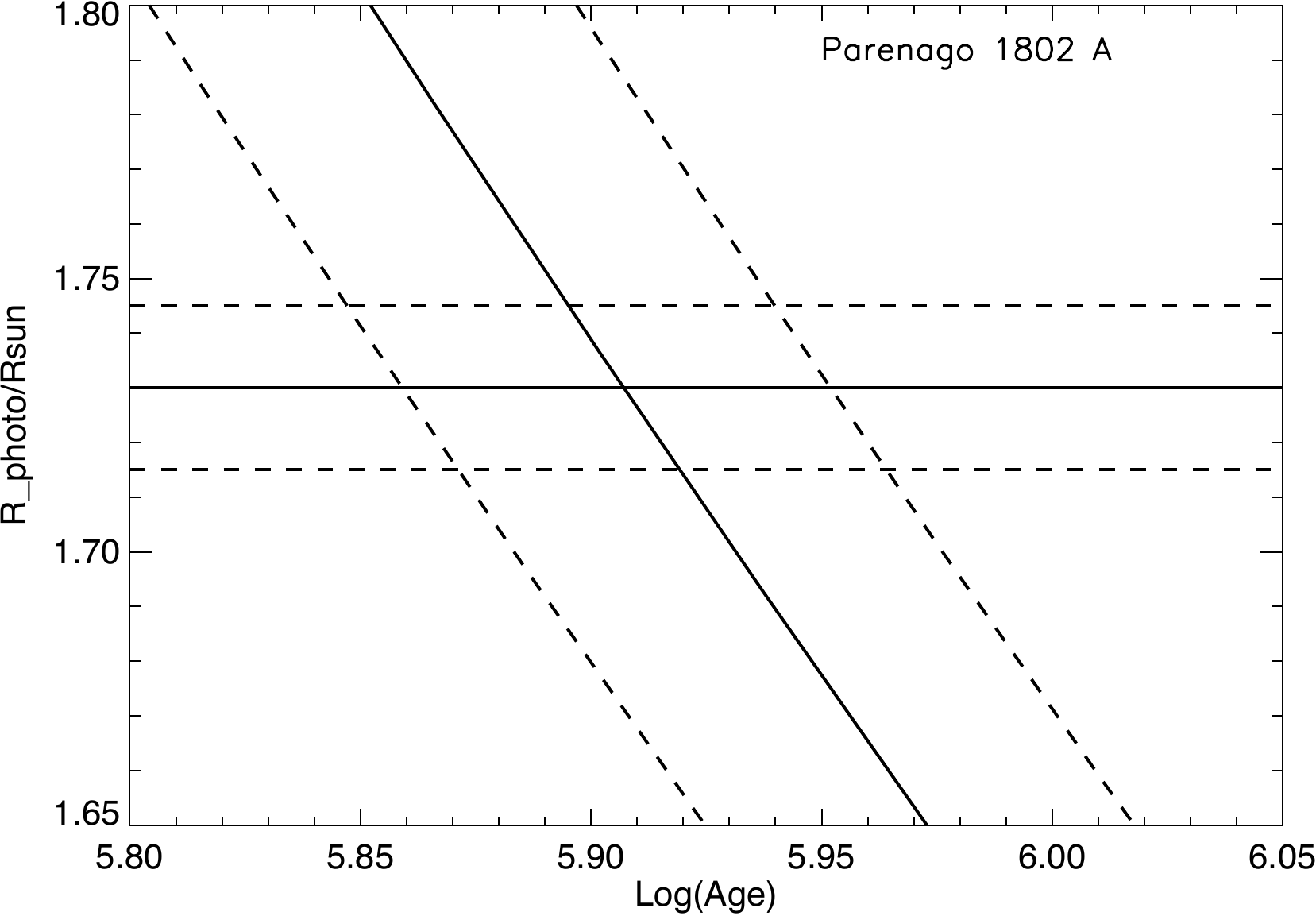}{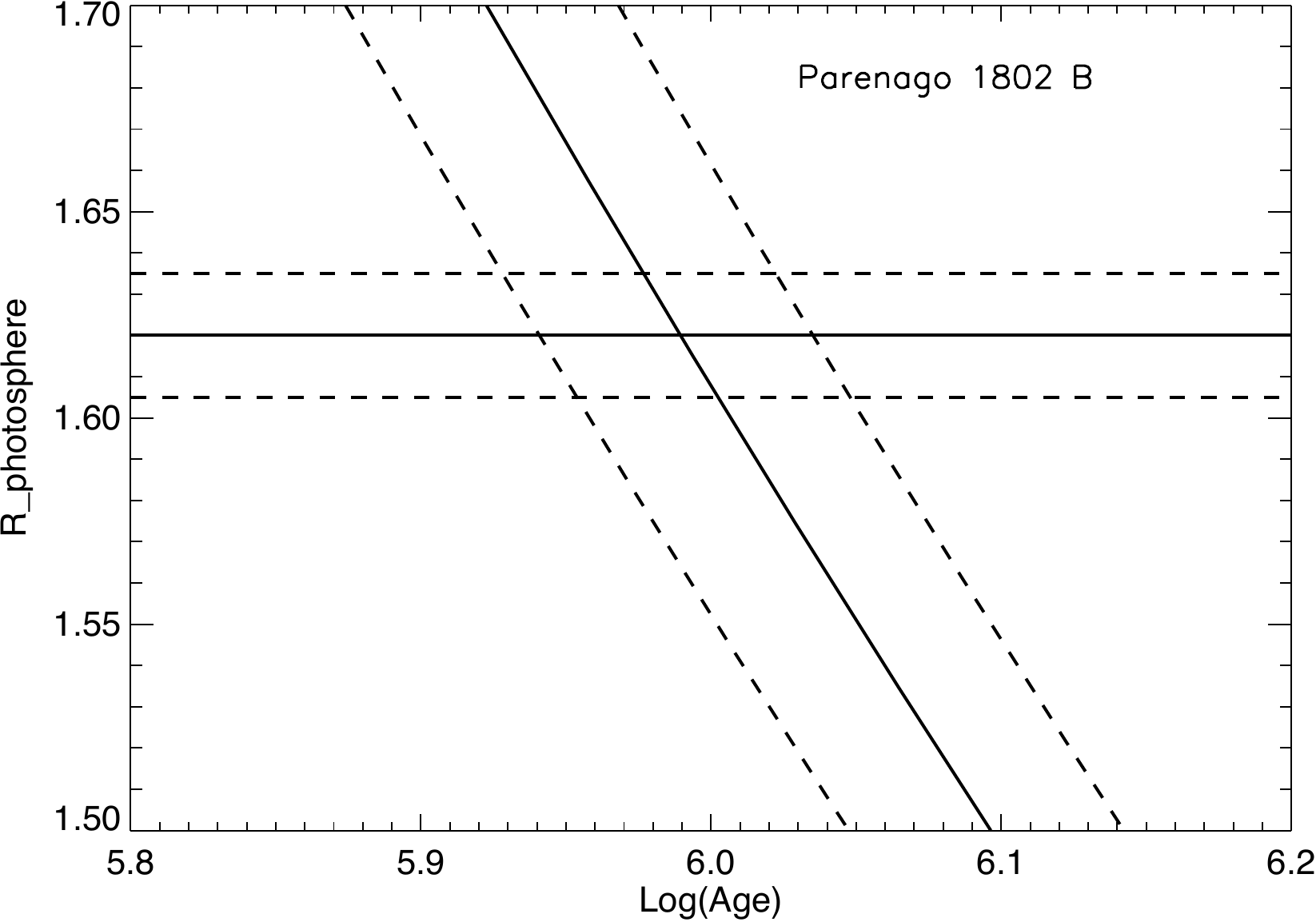}
\caption{Same as Fig. 2 but for Parenago 1802. The (A-B) 
age difference is $-0.16\pm 0.11$ MY. }
\end{figure}

\begin{figure}
\centering
\figurenum{7}
\plottwo{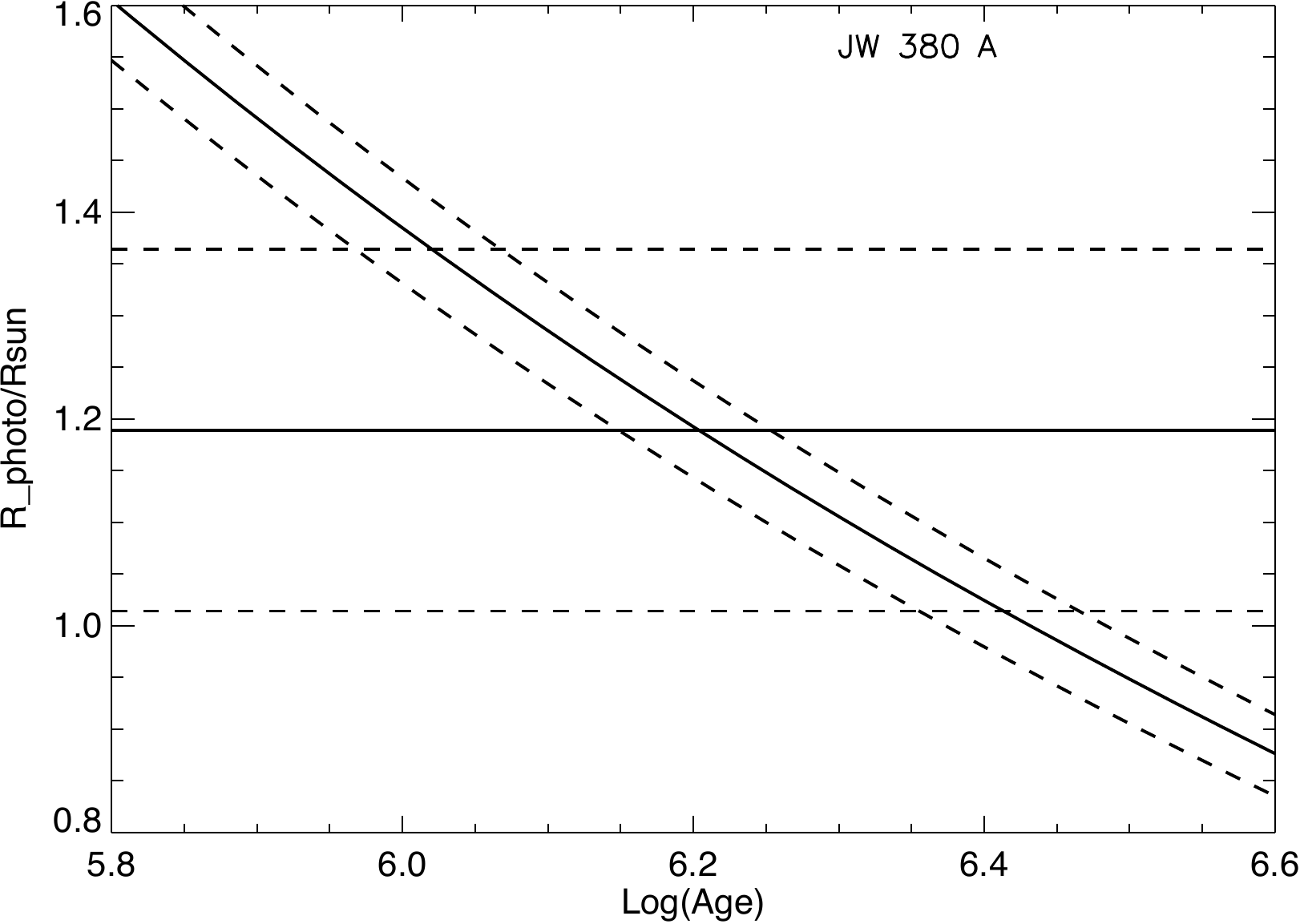}{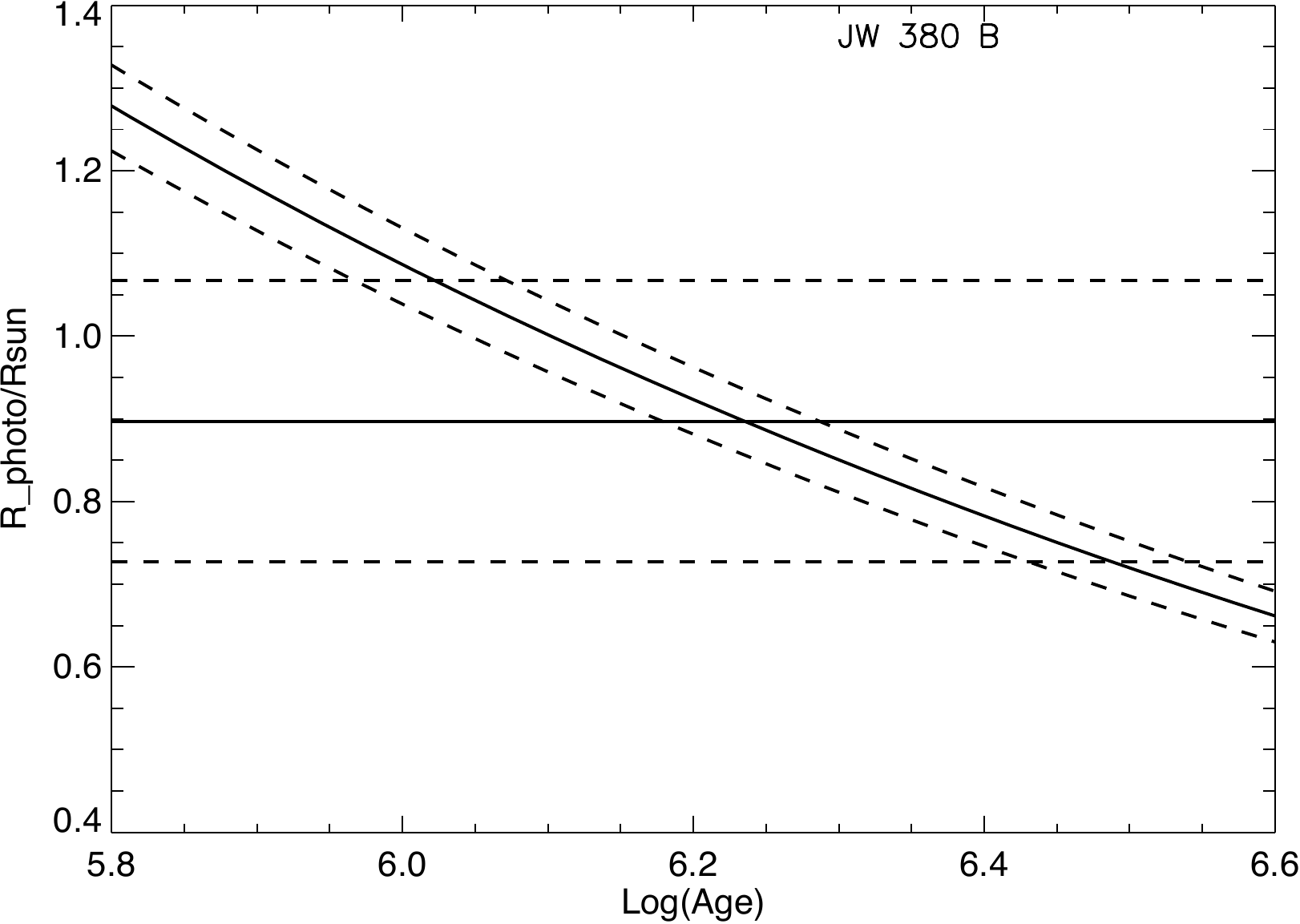}
\caption{Same as Fig. 2 but for JW380.  The (A-B) age difference is
$-0.09\pm 1.49$ MY.}
\end{figure}

\begin{figure}
\centering
\figurenum{8}
\plottwo{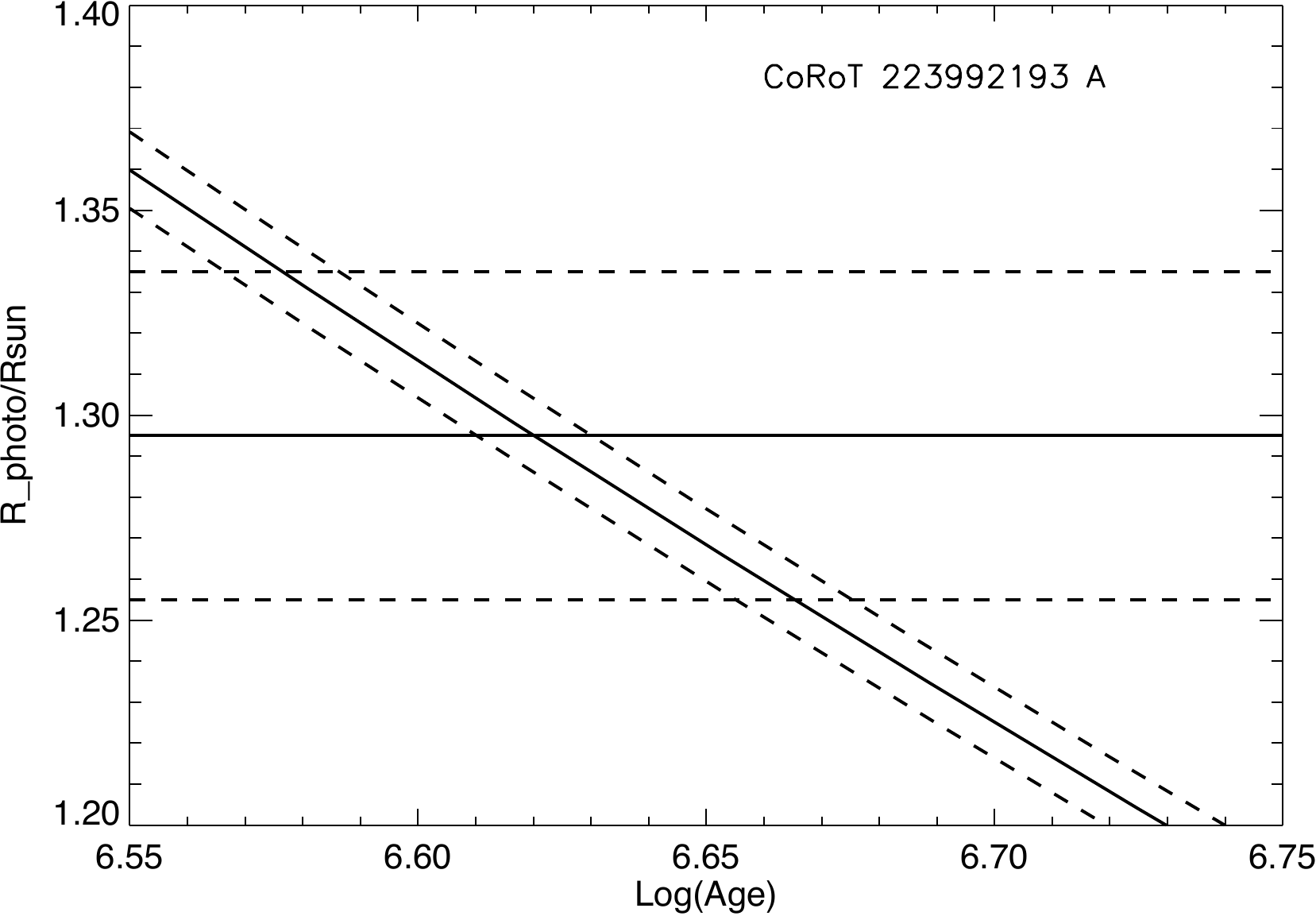}{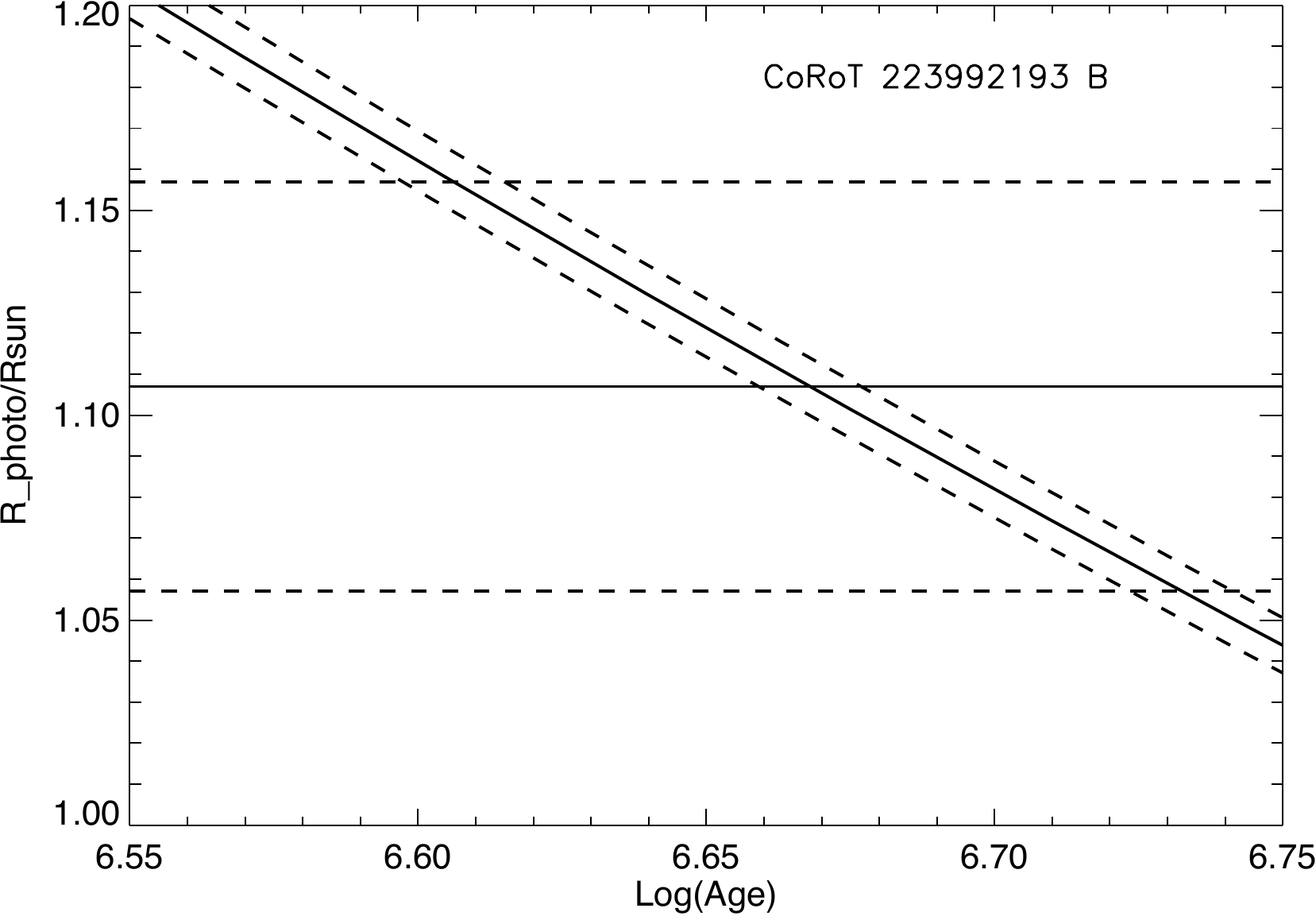}
\caption{Same as Fig. 2 but for CoRoT 223992193.   
The (A-B) age difference is $-0.51\pm 0.77$ MY. }
\end{figure}

\section{Discussion}

\subsection{V1174 Ori  A and B}

The position of V1174 Ori on the sky, its systemic velocity,
and small extinction suggest the EB is a member 
of the extensive Ori OB1d region 
and lies in front of Ori OB1c, the Orion Nebula Cluster
\citep{2004ApJS..151..357S}
 At face value the apparent age difference between 
the A and B components,~$\sim 2.7$MY~in a system of age 
$\sim 6$MY (Table 1), is enough to make the result suspect.
Could the age difference be  an unexpected artifact of the
our technique? Figure 5 shows that curvature of the 
$R_p~vs~log(Age)$~tracks at the masses and apparent
ages of the components is negligible. The mass of V1174 Ori B
is similar to that of CoRoT 223992195 A for which the analysis
(Fig. 8) does not indicate a problem at age similar to V1174Ori A.
The technique seems not to be at fault.

If the components really differ in age, where did they come from? 
Or, do their radii indicate differences in internal structure
possibly caused by tidal effects or heating of attributable
to the orbital evolution of an  as yet undetected tertiary as
\citet{2014NewAR..60....1S} suggested?  Or,  are
the measured parameters, the masses and radii, at fault and
produce the misleading result?  We consider these 
possibilities in turn.

As a member of Ori OB1d the EB's components must have formed when
the cluster was younger and more compact. Owing to a higher
stellar density, a young  binary or triple could have 
experienced the scattering events discussed by 
\citet{2012Natur.492..221R} and evolved into a hierarchical triple
consisting of the EB and  third member now in the aperture of the 
observations we have described.  This scenario is possible but
would not explain the large age difference.

\citet{2004ApJS..151..357S} suggested two dynamical effects
that could produced anomalous parameters of A and/or B:
tidal heating at the close orbital separations of the EB, and
energy input to the EB components by the orbital evolution
of a tertiary.  They presented a plausible scenario for the
tertiary's effects but neither this mechanism nor tidal heating
would explain why the effects appear in only one of the EBs
we analyzed.

Finally we consider the reliability of the masses and radii.
The spectral types of A and B are K4.5 and M1.5, respectively
\citep{2004ApJS..151..357S}.
The system's radial velocities yielded
velocity semi-amplitudes $K_{A,B}$ ~and hence
 $M_A sin^3 i = 1.004 \pm 0.016 \msun$~and
$M_B sin^3 i = 0.727 \pm 0.009\msun$. Using $P,e,K_A,
\&K_B$~ and an estimate of the masses from the 
spectral types we can make an independent 
estimate that the system's semi-major axis is 
$a <\sim 10 \rsun$  and $i>\sim 84^\circ$
~( \citet{2004ApJS..151..357S} determined that
the actual value is $86.97^\circ$).  If  
in fact $i=84^\circ$, the component masses would 
be only $\sim 2\%$ larger than the values
derived by \citet{2004ApJS..151..357S}. This is 
comparable to the uncertainties
of measurement of $K_A,K_B$.  It seems likely that
the component mass values are reliable and  not 
responsible for the calculated age difference. 

\citet{2004ApJS..151..357S} used the WD procedure to 
fit the light curve and thus to derive the component 
radii. It is beyond the scope of our work to estimate
the effects on the radii of the several assumptions
necessary to initiate and iterate the WD analysis.
We can imagine two approaches to assess whether the 
measured radii are reliable. One would be to try
to determine the origin of the ``third light''.
The other would be to analyze the V1174 Ori light 
curves independently with  procedures such as JKTEBOP.

\subsection{The Coeval EBs}

Five EBs in Table 1, ASAS J052821+0338.5, Parenago 1802,
JW 380, CoRoT 223992193, \& UScoCTIO 5, are truly coeval 
with an average age difference of only 
$\sim 0.3$ MY. Their  ages as indicated
by the {\it MESA} models lie in the range $\sim
1 - 6$ MY. These results are robust because
  Fig. 1 indicates that their apparent 
ages would be similar if they 
were measured using the BHAC15 and F16 models.
Since the slopes of  the three models are nearly 
identical for the masses and ages considered,
the diferences in ages would be nearly identical too.    
The precision of the derived
ages, and hence also the difference in component
age, is limited only by the precision of the 
measured radii and the slope of the decrease of 
stellar radius with time.
These results indicate that the accretion events 
and early orbital evolution of the compnents in 
these EBs produced and age spread of no more 
than 0.3 MY.

The rate of contraction of the PMS star is determined by
gravity and its internal pressure.  In the presence
of both gas pressure and magnetic pressure the contraction
is slowed (F16; \citet{2017ApJ...834...67M}.
Thus at a given age the radius is greater than
the case without an internal magnetic field.
Fig. 9  shows that the slopes of the magnetic models (F16) 
and {\it MESA} models, which are non-magnetic, are similar
for ages 1 - 6.3 MY.  Application of the magnetic 
models to our sample would yield age estimates that are
younger but with component age differences that are
consistent with the values in Table 1. Thus, the ages 
given in Table 1 are precise but not necessarily accurate.
However, the component age differences are accurate and precise.

\begin{figure}
\centering
\figurenum{9}
\includegraphics[width=7cm]{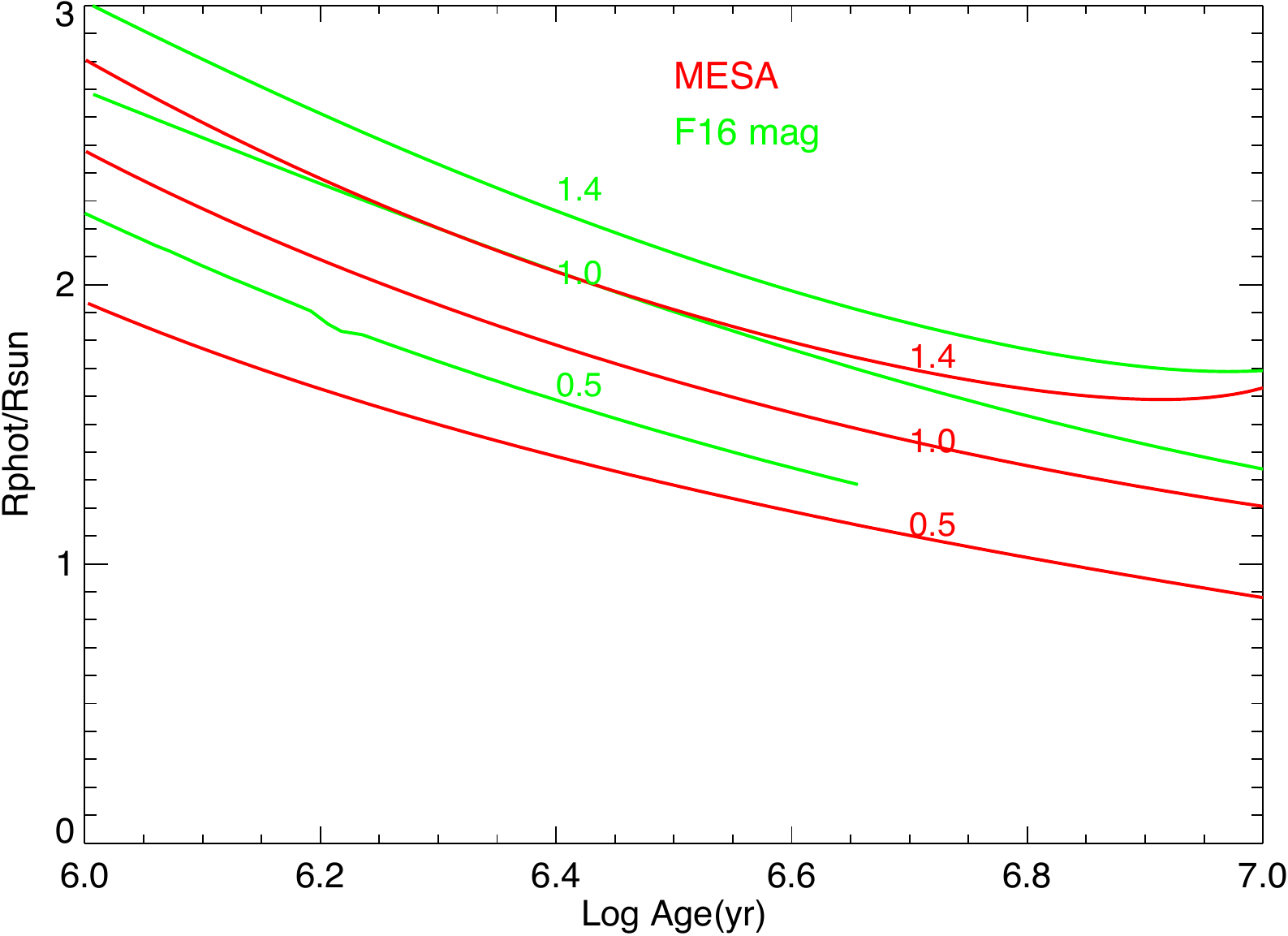}
\caption{A comparison of the $R~vs~log(Age)$ 
evolutionary tracks of {\it MESA} models and  F16 models that 
include interior magnetic fields. }
\end{figure}

 ASAS J05281+0338.5, Parenago 1802, and JW 380 have 
mass ratios close to 1.  The stars in Parenago 1802 can be 
considered ``twins'' in the sense that their mass ratio 
is  within $2\%$ of 1.0 \citep{2009AJ....137.3442S}. Evidently
the processes that produce EBs within a few tenths MY
can result in EBs with  very nearly equal component masses.

EBs that are actually triples can provide
more information on their architectures and hence formation.
The triples are presumably stable on at least 1-10 MY
time scales with orbital separation of the tertiary 
much greater than that of  the EB.  \citet{1995ApJ...455..640E}'s
stability criterion provides an estimate of the 
distance at which the orbit of tertiary is stable.  They 
find that, in terms of the ratio of the periastron 
distance of the tertiary to the apostron distance of 
the inner binary  

$$ Y^{min} = {{a_{outer}(1-e_{outer})}\over{a_{inner}(1+e_{inner})}}  $$

\noindent
{the triple is stable when $Y^{min} > Y^{min}_0$. For
equal masses $Y^{min}_0\sim 5$.  Thus, if Parenago 1802 bears
out the suspicion provided by its third light and proves
to be a triple $ a_{outer} \ge 0.25$AU in which the lower
bound would be realized if $e_{outer}=0$. At Parenago 1802's 
distance in the Orion Nebula Cloud, $\sim 400$pc
(e.g. \citet{2007MNRAS.376.1109J} the separation of the EB 
and tertiary is  $0.0006''$~ if $e_{outer}=0$~and greater 
otherwise.  This separation is unfortunately unresolvable 
by the largest aperture ground-based telescopes.

The high precision of the age differences of these EBs 
is attributable to the very high precision and model independent
measurements of their masses and radii.  The traditional technique 
of age determination of EBs by fitting observations to  model isochrones
on the HRD requires the luminosity and effective
temperature.  Both are derived quantities and inevitably have
uncertainties larger than that of the radii alone. 
It seems very unlikely that the accuracy and precision
component age differences we have achieved could be 
reached using HRD techniques. 

\section{Summary and Suggestions for Future Work}

There are two main results of our work:
\vskip 0.5cm

1)  Using as inputs only the precision masses and radii of the
components of PMS EBs, evolutionary tracks calculated using the
{\it MESA} software provide accurate and precise values of their
age differences.

2)  All 5 of the EBs that we could analyze with confidence,
 ASAS J052821+0338.5, Parenago 1802, JW 380, CoRoT 223992193, 
and  UScoCTIO 5, are coeval to 0.3 MY on average,  This means
that the complex events such as accretion and orbital evolution
associated with the formation of very close binaries produced
an apparent age spread of no more than 0.3 MY.

\vskip 0.5cm
The following observational advances of multiples
such as those we studied would contribute significantly to an 
understanding of their formation:

1) Further study of the systems with ``third light'' would enable 
assessment of its physical origin.

2) Suspected circumbinary disks such as in CoRoT 223992193  
\citep{2014A&A...562A..50G} could be revealed  
by radio interferometric observations and would open a 
new channel for the study of EB formation.

3) Analysis of and EB alone does not provide a direct measurent
of its distance; instead it is determined from the
luminosities and effective temperatures of its components
and reference to models of stellar evolution.  {\it GAIA} 
parallaxes
with microarc second uncertainties will provide distances with
unsurpassed precision.

\acknowledgements

We are truly grateful to the referee for an exceptionally helpful
report that helped us clarify the presentation.
We thank K. Stassun, G. Torres, and G. Feiden for
comments that improved an earlier draft of this paper.
We thank I. Baraffe, G. Feiden, B. Paxton, A. 
Dotter, and J. Schwab for very helpful advice about  
calculation of theoretical models of PMS evolution.

\appendix

\section{MESA information}

We used MESA version 7624 to calculate our models.  
We used metallicity z=0.02
and the default abundances.  We set the mixing length 
parameter to 1.918 which 
reproduces the Sun at present age.  When this paper 
is accepted we will 
provide sample inlists at http://mesastar.org/results.

\clearpage

\begin{table}
\caption{Pre-Main Sequence Eclipsing Binaries Considered}
\begin{tabular}{lc|cll|lr}
\hline
\hline
     &            &\multicolumn{3}{c}{Derived from Observations}&\multicolumn{2}{c}{This Work}\\
  Name/Ref.  & Comp.  & P(d) &    $M_*/\msun$       & $R_{p}/\rsun$ & log Age(y)    & Age (A-B) MY\\
\hline
ASAS J05281+0338.5&A&3.87 &$1.375\pm0.028$  &$1.83\pm0.07$&$6.546\pm0.066$&  \\
1,2                 &B&   &$1.329\pm0.020$  &$1.73\pm0.07$&$6.607\pm0.070$&$ -0.53\pm0.85$\\
RX J0529.4+0041   &A&3.04 &$1.27\pm0.01$   &$1.44\pm0.10$&$N/A$          &   \\
1,3                 &B&   &$0.93\pm0.01$   &$1.35\pm0.10$&$6.741\pm0.12$&   \\
V1174 Ori         &A&2.64  &$1.006\pm0.013$&$1.338\pm0.011$&$6.809\pm0.012$& \\
1,4              &B&       &$0.7271\pm0.0096$   &$1.063\pm0.011$&$6.962\pm0.016$& -$2.72\pm0.39$\\
Parenago 1802   &A&4.67  &$0.391\pm0.032$&$1.73\pm0.015$&$5.904\pm0.040$&   \\
1,5              &B&       &$0.385\pm0.032$      &$1.62\pm0.015$&$5.982\pm0.040$&-$0.16\pm0.11$\\
JW 380            &A&5.30  &$0.262\pm0.025$&$1.189\pm0.175$ &$6.200\pm0.220$&\\
1,6             &B&       &$0.151\pm0.013$      &$0.897\pm0.170$ &$6.225\pm0.214$ &$-0.09\pm1.49$\\
CoRoT 223992193  &A&3.88 &$0.668\pm0.012$  &$1.295\pm0.040$&$6.622\pm0.044$& \\
1,7              &B&      &$0.4953\pm0.0073$      &$1.107\pm0.050$&$6.672\pm0.063$&$-0.51\pm0.77$\\
USco CTIO5&  A   & 34.0 &$0.329\pm 0.002$& $0.834\pm0.006$&$6.798\pm0.010$& \\
8         &  B     &        &$0.317\pm 0.002$& $0.810\pm0.006$&$6.811\pm0.010$&$-0.19\pm0.22$\\
USco CTIO5&  A  & 34.0 &$0.3336\pm 0.0022$& $0.862\pm0.012$&$6.763\pm0.018$& \\
9         &  B  &      &$0.3200\pm 0.0022$& $0.852\pm0.013$&$6.754\pm0.019$&$0.12\pm0.34$\\
\hline
\end{tabular}
\end{table}
\parindent=0.0in
References: 1) Stassun et al. (2014), 2) Stempels et al. (2008), 3) Covino et al. (2004),
4) Stassun et al. (2004), 5) Gomez Maqueo Chew et al. (2012), 7) Gillen et al. (2007), 
8) Kraus et al. (2014), 9) David et al (2014)

\end{document}